\documentclass{ws-ijmpcs}

\def\be{\begin{equation}}
\def\ee{\end{equation}}
\def\bea{\begin{eqnarray}}
\def\eea{\end{eqnarray}}

\def\gev{\, {\rm GeV}}

\newcommand{\sigmaSI}{\sigma_{\rm SI}}

\newcommand{\gsim}{\lower.7ex\hbox{$\;\stackrel{\textstyle>}{\sim}\;$}}
\newcommand{\lsim}{\lower.7ex\hbox{$\;\stackrel{\textstyle<}{\sim}\;$}}

\newcommand{\pb}{\rm pb}

\newcommand{\cm}{\rm cm}

\begin{document}

\markboth{Jason Kumar}
{Probing Isospin-Violating Dark Matter}

%
\catchline{}{}{}{}{}
%

\title{Probing Isospin-Violating Dark Matter}

\author{Jason Kumar}

\address{Department of Physics and Astronomy, University of Hawaii,
2505 Correa Road\\
Honolulu, Hawaii 96822,
United States of America
\\
jkumar@hawaii.edu}

\maketitle

\begin{history}
\received{30 December 2011}
\end{history}

\begin{abstract}
We discuss experimental probes of isospin-violating dark matter (IVDM),
including direct and indirect detection strategies.  We point out
the important role which IVDM plays in understanding recent data regarding
low-mass dark matter, and describe strategies for finding evidence of
IVDM at current and upcoming experiments.

\end{abstract}

\ccode{PACS numbers: 11.25.Hf, 123.1K}

\section{Introduction}	

Dark matter search strategies at direct detection experiments are typically based on
several nominal assumptions.  Among these assumptions are
\begin{itemize}
\item{...that dark matter has a Maxwellian velocity distribution, and a density
$\sim 0.3~\gev / \cm^3$ near the earth.}
\item{...that dark matter scatters elastically off nuclei.}
\item{...that dark matter interacts with Standard Model matter through
effective contact operators (equivalently, the mediating particles are heavy).}
\item{...that dark matter interactions are isospin-invariant (that is, that
dark matter interacts with protons and neutrons in the same way).}
\end{itemize}
One must keep in mind that any deviation from these assumptions
can modify the comparison of experimental results from different
experiments.

In these proceedings, we focus on the last assumption: isospin-invariant
interactions.\cite{IVDM,Chang:2010yk,arXiv:1102.4331}
The main motivation for this assumption is the fact that, in most models where the
dark matter is the lightest neutralino of the MSSM, dark matter interactions are indeed largely
isospin-invariant.  However, this motivation loses force as soon as one considers
models beyond LSP WIMPs.

Isospin-violating dark matter (IVDM) can have an important impact on the interpretation
of data.  This is especially relevant when considering low-mass dark matter ($m_X \sim 5-20~\gev$).
DAMA,\cite{Bernabei:2010mq} CoGeNT\cite{Aalseth:2010vx,Aalseth:2011wp}
and CRESST\cite{Angloher:2011uu} have reported
data consistent with low-mass dark matter, though not necessarily consistent with
each other.  However, CDMS,\cite{Akerib:2010pv,Ahmed:2010wy}
XENON10/100\cite{Angle:2011th,Aprile:2011hi,Aprile:2010um}
and SIMPLE\cite{Felizardo:2011uw} have
reported data which is in tension with this interpretation.  Issues have been
raised regarding the credibility of all these experimental results, and
we will not delve into these issues any further.  Instead, we will focus on the fact
that the tension between these data sets has been found under the assumption of
isospin-invariant interactions; this situation changes significantly for IVDM.  We will see
that neutrino-based indirect detection searches have an interesting
complementary role in probing IVDM.

\section{Isospin-Violating Dark Matter}

For simplicity, we assume dark matter scatters elastically
with Standard Model nuclei through contact interactions.  In this case,
the rate at which dark matter-nucleus scattering occurs can be written as
\begin{equation}
R =N_T n_X  \! \! \int \!  d E_R \!
\int_{v_{\text{min}}}^{v_{\text{max}}} \! \! \!  d^3 v \, f(v) v
\frac{d\sigma}{dE_R} \ ,
\end{equation}
where $N_T$ is the number of target nuclei, $n_X$ is the dark matter number
density near the earth, $\sigma$ is the dark matter-nucleus scattering
cross-section and $E_R$ is the nuclear recoil energy.  $f(v)$ is the dark matter
velocity distribution and $v_{min}=\sqrt{m_A E_R / 2\mu_A^2}$ is the minimum
dark matter velocity for which it is kinematically possible to produce a recoil
energy $E_R$.  The dark matter-nucleus reduced mass is $\mu_A = m_A m_X /(m_A+ m_X)$.
$v_{max}$ is determined by the galactic escape velocity, and the $E_R$ integration limits
are determined by the thresholds of the experiment.

The differential cross-section can be written as
\bea
{d\sigma \over dE_R} &=& {m_A \over 2v^2 \mu_A^2} \hat \sigma_A ,
\eea
where
\bea
\hat \sigma_A &=& {\mu_A^2 \over M_*^4} [f_p Z F_A^p (E_R) + f_n (A-Z) F_A^n (E_R)]^2.
\eea
$M_*$ is an overall energy scale, $f_{p,n}$ are the
dark matter couplings to protons and neutrons, and $F_A^{p,n}(E_R)$ are nuclear form factors.
Isospin-invariant interactions imply $f_n = f_p$.
The nuclear form factors for protons and neutrons are not
the same, and thus are an additional source for isospin-violating interactions.  However, as
this effect is relatively small compared to the effect of $f_n \neq f_p$, will assume
(for simplicity) $F_A^{p}(E_R) \sim F_A^{n}(E_R) = F_A (E_R)$.  We can then
write $R = \sigma_A I_A$, where
\bea
\label{SIcross}
\sigma_A &=& \frac{\mu_A^2}{M_*^4}
\left[ f_p Z + f_n (A-Z) \right]^2 ,
\\
I_A &=&N_T n_X\! \! \int \!  d E_R \!
\int_{v_{\text{min}}}^{v_{\text{max}}} \! \! \!  d^3 v \,
f(v)  \frac{m_A}{2 v \mu_A^2} F_A^2 (E_R) \, .
\eea
$I_A$ encodes the nuclear and astrophysics, while $\sigma_A$ encodes the dark matter
particle physics; it is the latter factor we will focus on.

The cross-section for dark matter to scatter off a single proton, $\sigmaSI^p$, is then
\bea
\sigmaSI^p = {\mu_p^2 f_p^2 \over M_*^4},
\eea
in terms of which the event rate can be written as
\bea
R = \sigmaSI^p \sum_i \eta_i \frac{\mu_{A_i}^2}{\mu_p^2}
I_{A_i} \left[Z + (A_i-Z) f_n / f_p \right]^2 .
\eea
This allows one to relate an observed event rate to $\sigmaSI^p$ for
a given value of $f_n / f_p$.
Here, $\eta_i$ is the natural abundance of each isotope $i$, with atomic
mass $m_{A_i}$.

Dark matter signal and exclusion regions
are typically expressed in terms of $\sigma_N^Z$, the normalized dark matter-nucleon
scattering cross-section.
$\sigma_N^Z$ can be determined from the rate
above by setting $f_n = f_p$, and is related to $\sigmaSI^p$ by
\bea
F_Z \equiv \frac{\sigmaSI^p}{\sigma_N^Z}
= \frac{\sum_i \eta_i \mu_{A_i}^2 A_i^2}
{\sum_i \eta_i \mu_{A_i}^2 [Z + (A_i - Z) f_n/f_p]^2}.
\eea
$F_Z$ depends only on known atomic physics, and on $f_n / f_p$.
If $f_n \neq f_p$,
the dark matter-nucleon scattering cross-section is not really a sensible physical
quantity.  $F_Z$ represents the factor by which the normalized-to-nucleon cross-section
reported by an experiment (assuming isospin-invariant interactions) must be scaled to
obtain the physical cross-section for dark matter scattering off a proton.

\section{Low-mass dark matter}
\begin{figure}[tb]
\includegraphics*[width=0.49\columnwidth]{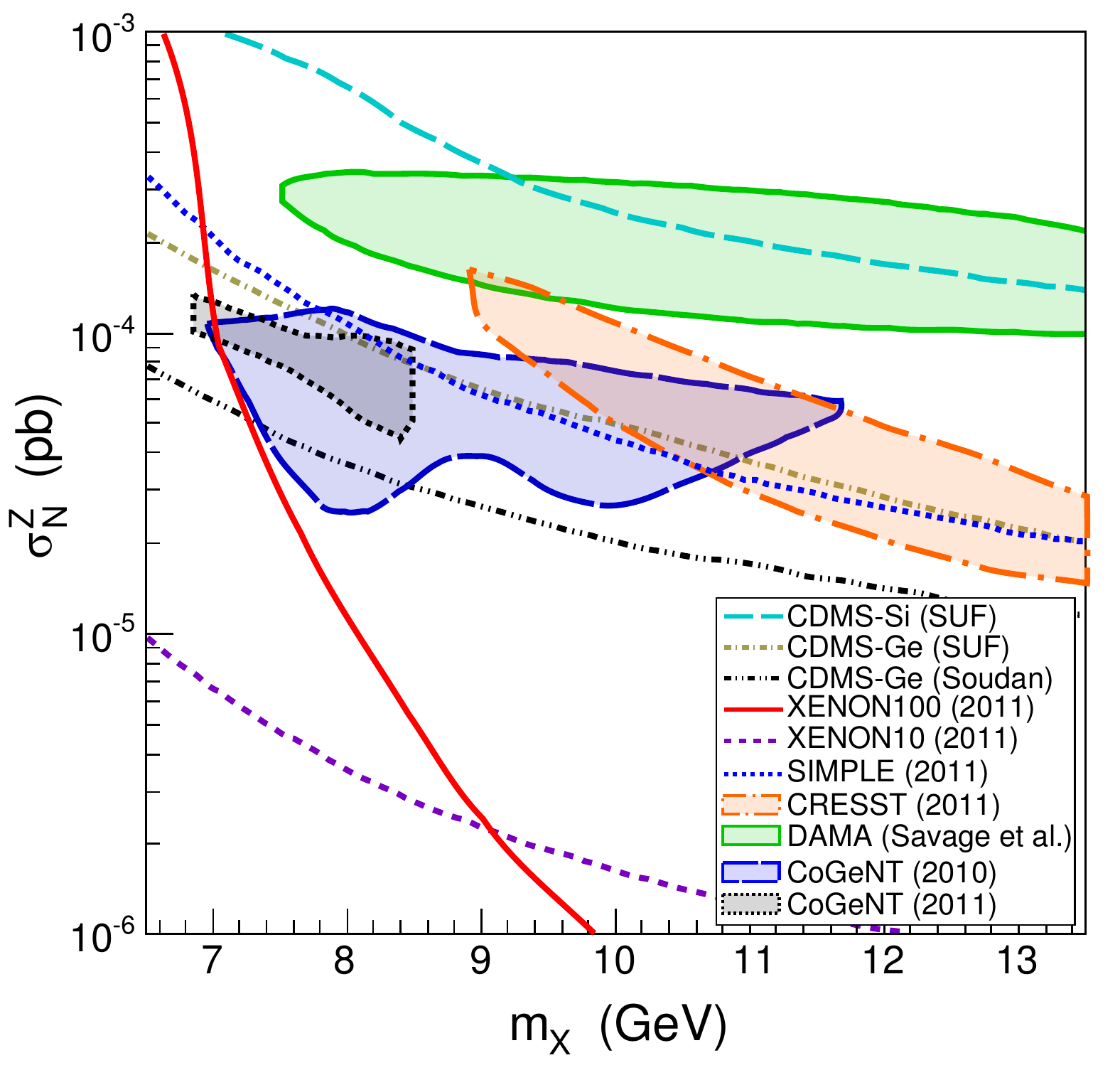}
\includegraphics*[width=0.49\columnwidth]{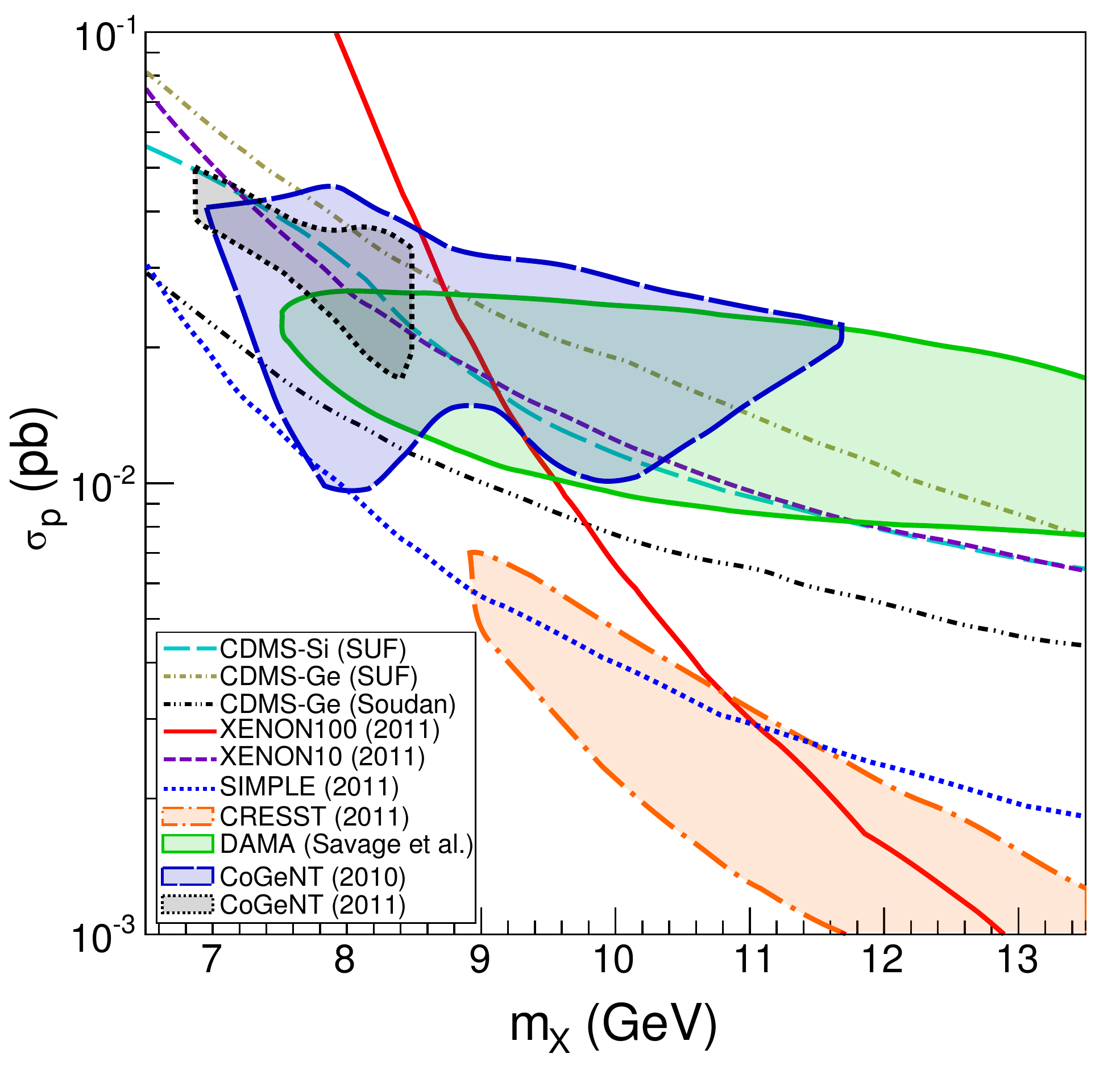}
\vspace*{-.1in}
\caption{\label{fig:DDbounds} Favored regions and exclusion contours
in the $(m_X, \sigma_N^Z)$ plane (left),  and in the
$(m_X, \sigma_p)$ plane for IVDM with $f_n / f_p = - 0.7$ (right).
(Figure courtesy of David Sanford.) }
\end{figure}

We now apply this analysis to the low-mass dark matter data.  In fig.~\ref{fig:DDbounds}
(see ref.~\refcite{arXiv:1102.4331}),
we plot signal regions for DAMA\cite{DAMAregion} ($3\sigma$),
CoGeNT\cite{Aalseth:2011wp} (90\% CL) and CRESST\cite{Angloher:2011uu} ($2\sigma$),
as well as 90\% CL exclusion contours for CDMS,\cite{Akerib:2010pv,Ahmed:2010wy} XENON10,\cite{Angle:2011th}
XENON100\cite{Aprile:2011hi} and SIMPLE\cite{Felizardo:2011uw}.
The left panel assumes $f_n / f_p =1$, while
the right panel assumes $f_n / f_p =-0.7$.  This latter value yields
the maximum suppression possible for the dark matter-xenon scattering cross-section
due to interference between protons and neutrons.
Isospin-violating interactions have a dramatic effect on the consistency of data sets
from different experiments.\cite{Chang:2010yk,arXiv:1102.4331,arXiv:1110.4616}  For
$f_n / f_p \sim -0.7$, the DAMA and CoGeNT signal regions are consistent
($m_X \sim 8~\gev$, $\sigmaSI^p \sim 3\times 10^{-2}~\pb$) and satisfy bounds
from xenon-based experiments, which are the most constraining bounds for
isospin-invariant interactions.

On the other hand, IVDM may not provide a
complete reconciliation of all the data.\cite{arXiv:1110.2721,arXiv:1110.5338}  There is marginal tension between
the exclusion contour of CDMS (Soudan) and the signal region of CoGeNT; this
cannot be alleviated by isospin-violating interactions, as both data sets use
germanium detectors.  Moreover, the bounds from SIMPLE are more constraining
on models which can match the CoGeNT data if $f_n / f_p \sim -0.7$.
Even the signal regions cannot be brought into perfect alignment; the choice of
$f_n / f_p$ which reconciles the DAMA and CoGeNT data does not reconcile the
CRESST data.

The experimental situation may change significantly in the near future.
CoGeNT has indicated that their experiment may have more
surface area contamination than previously thought.\cite{CollarTaup2011}  Preliminary indications
are that this would move their signal region to larger mass and smaller
$\sigmaSI^p$, bringing it more in line with CRESST.  Understanding of
xenon's response to low-energy recoils is steadily improving.
Many experimental uncertainties may be clarified with more data.

More of the data may be brought into
alignment if, in addition to isospin-violating interactions, one weakens other
assumptions\cite{arXiv:1105.3734,arXiv:1108.1391,arXiv:1109.4639}
by allowing low-mass mediated interactions, inelastic scattering, and/or non-Maxwellian
velocity distributions.  It is clear, however, that isospin-violating
interactions can have a large effect on the understanding of the low-mass data, and
must be taken into account.\cite{arXiv:1110.5338}

\section{New Experiments}

For $f_n / f_p \sim -0.7$, the sensitivity of CRESST and SIMPLE
increased relative to CoGeNT.  This is because they have a carbon, oxygen and fluorine
target nuclei, for which the neutron-to-proton ratio is smaller than that of germanium.
Several other experiments with similar targets can test these low-mass
IVDM models.  These include COUPP,  DMTPC, and the Directional Dark matter Detector (${\rm D^3}$),
whose prototype is in
the commissioning phase at the University of Hawaii.\cite{Vahsen:2011qx}  One can parameterize the sensitivities they
can achieve for $f_n / f_p \sim -0.7$ by computing the ratio of the normalized-to-nucleon
cross-section inferred from these target nuclei to that inferred from germanium:
\bea
\sigma_N^{Z=C} &\sim& 8.4 \times \sigma_N^{Z=Ge},
\nonumber\\
\sigma_N^{Z=O} &\sim& 8.5 \times \sigma_N^{Z=Ge},
\nonumber\\
\sigma_N^{Z=F} &\sim& 4.2 \times \sigma_N^{Z=Ge}.
\eea

Given the confusing experimental situation regarding direct detection data,
it is worthwhile to consider alternative tests of IVDM.
An interesting method utilizes neutrino detectors, which search
for the flux of neutrinos arising from dark matter annihilation in the core of the sun.  The neutrino
flux is determined by the dark matter annihilation rate.  Assuming that
the sun is in equilibrium, the dark matter annihilation rate is half of the capture rate, which is
largely determined by $\sigmaSI$.  The sun contains many low-neutron targets which are less susceptible
to destructive interference between proton and neutron couplings (for hydrogen,
there is no destructive interference).  Super-Kamiokande is capable of providing
competitive sensitivity to low-mass dark matter,\cite{SKlowmass}
and its sensitivity to IVDM allows it to probe
models which could match the low-mass data.\cite{arXiv:1106.4044}

\begin{figure}[tb]
\includegraphics[width=0.49\columnwidth]{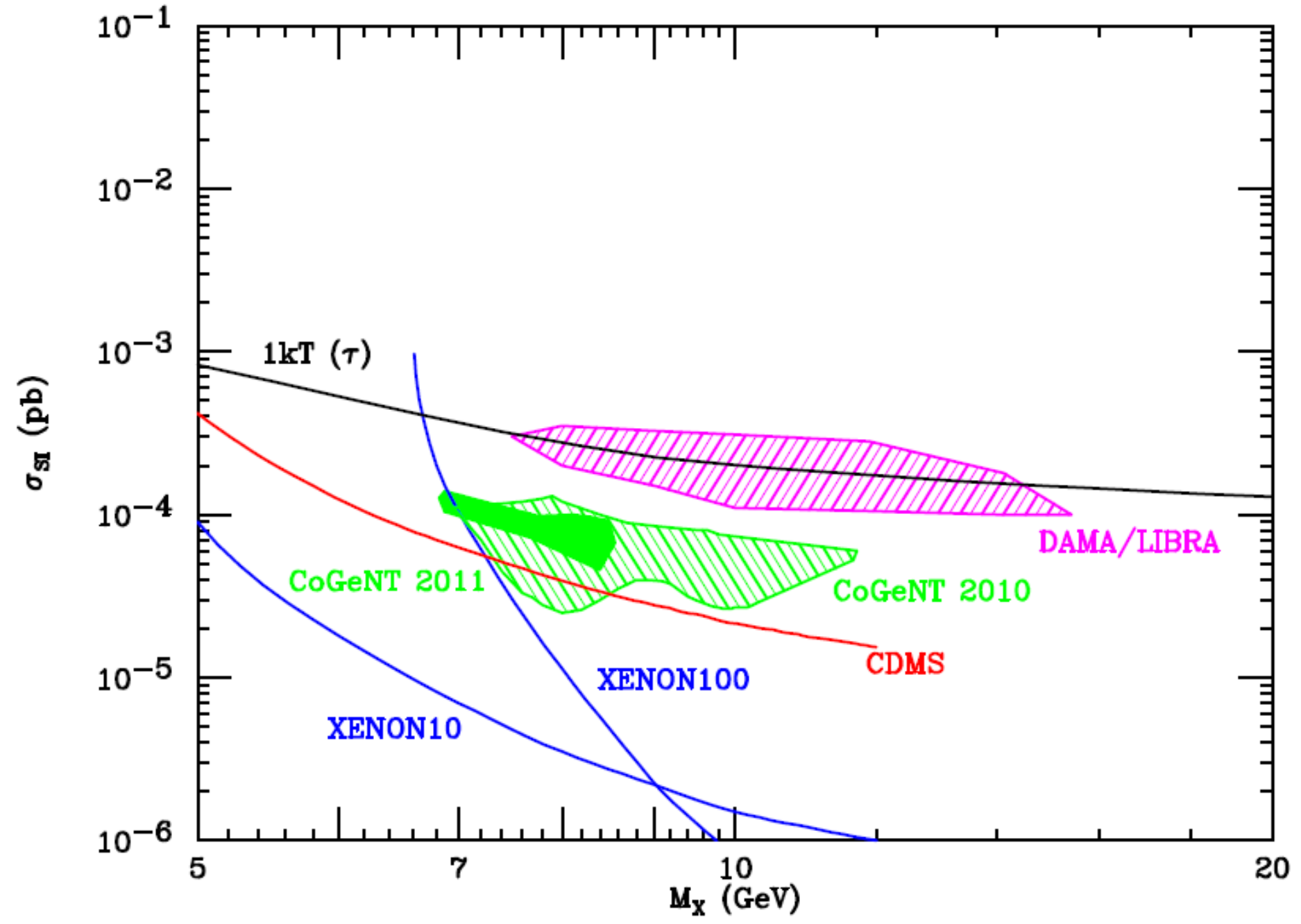}
\includegraphics[width=0.49\columnwidth]{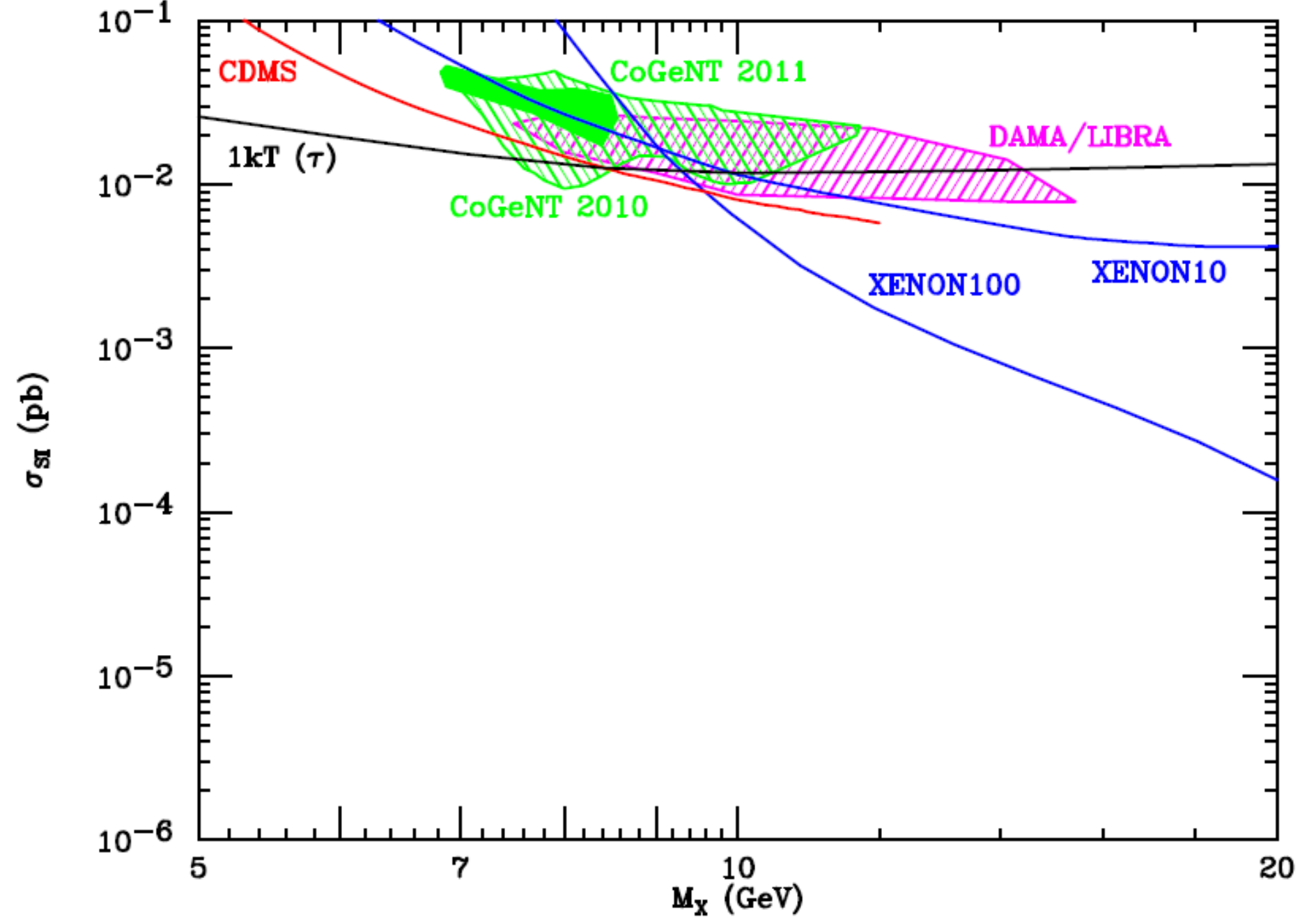}
\vspace*{-.1in}
\caption{
\label{fig:SIbounds}  Sensitivity of a 1 kT LS detector (such as KamLAND)
to $\sigmaSI$, assuming 2135 live-days and annihilation to $\tau \bar \tau$.
Also plotted are signal regions and exclusion contours for
other experiments (see text).  The left panel assumes $f_n = f_p$.
The right panel is for $f_n / f_p = -0.7$, conservatively assuming
dark matter capture only from scattering off H.  (Figure courtesy
of Stefanie Smith.)
}
\end{figure}

Liquid scintillation neutrino detectors, such as KamLAND can also be sensitive to IVDM models.\cite{arXiv:1103.3270}
In this analysis, one must rely on the ability to reconstruct the track of the lepton
produced in a charged-current interaction from the timing of when the first scintillation photons
reach the photomultiplier tubes.\cite{LSdet}  This reconstruction can allow one to determine
the direction and energy of a fully-contained charged lepton, allowing one to determine if
the initial neutrino came from the direction of the sun.  Moreover, it permits lepton
flavor discrimination, which is useful for selecting events produced by electron neutrinos.  An analysis
using electron neutrinos has the advantage of a much smaller atmospheric neutrino background.
Also, the quick attenuation of the electron shower allows one to measure the full energy of
the charged lepton.  Assuming the energy and angular resolutions found
in Ref.~\refcite{LSdet}, we plot in fig.~\ref{fig:SIbounds} (see ref.~\refcite{arXiv:1103.3270})
the sensitivity of a  1 kT LS detector operating for 2135 live-days and assuming
$f_n / f_p = 1,-0.7$.  Also plotted for reference are the signal regions of CoGeNT\cite{Aalseth:2010vx,Aalseth:2011wp}
and DAMA,\cite{DAMAregion} and exclusion
contours from CDMS\cite{Akerib:2010pv} and XENON10/100.\cite{Angle:2011th,Aprile:2010um}
KamLAND can potentially be sensitive to IVDM models (with
$f_n / f_p \sim -0.7$) which could match the low-mass data.

\section{Matching Multiple Experiments}

Moving beyond the low-mass data, there are two basic questions one can ask:
\begin{itemize}
\item{Given a signal at one detector, what is the minimum exposure
a different detector might need to confirm it, allowing for IVDM?}
\item{Given a signal at one detector, what is the maximum exposure
a different detector would need to definitely contradict it, even allowing for IVDM?}
\end{itemize}
We can answer this by determining the sensitivity to $\sigma_N^Z$, the normalized-to-nucleon
cross-section assuming isospin-invariant interactions, a second experiment would need to
either possibly confirm or definitely refute a signal from the first experiment.  These are given by
the maximum and minimum (with respect to $f_n / f_p$) of the ratio
\bea
R [Z_1 , Z_2](f_n / f_p) &\equiv& {\sigma_N^{Z_1} \over \sigma_N^{Z_2}} = {F_{Z_2} \over F_{Z_1}}.
\eea

The isotope content plays a key role in this analysis.  For an element
$Z$ with only one isotope, the quantity $R_{max}[Z_1 , Z]$ is infinite, allowing an IVDM model
with $f_n / f_p = Z/(Z-A)$ to evade the bounds from a $Z$-based detector.
This is not the case if there are multiple isotopes with a non-negligible
abundance; no choice of $f_n / f_p$ can cancel the response from all isotopes.
For an IVDM signal
at a Ge-based detector to be definitively probed by a Xe-based detector, the latter must have at
most a factor of $\sim 22$ greater sensitivity than the former.
Considering the great sensitivity of large Xe-based detectors, IVDM models
matching the low-mass data of CoGeNT, DAMA and
CRESST can eventually be probed at xenon detectors.

\section{IceCube/Deepcore}

We have seen that the sensitivity of neutrino detectors to IVDM can be significantly enhanced.
For large $m_X$, IceCube/DeepCore has the greatest sensitivity of all neutrino detectors.
We compare its sensitivity to IVDM to that of other current and future direct
detection experiments, for a range of masses and values of $f_n / f_p$.\cite{arXiv:1108.0518}

\begin{figure}[tb]
\includegraphics[scale=0.32]{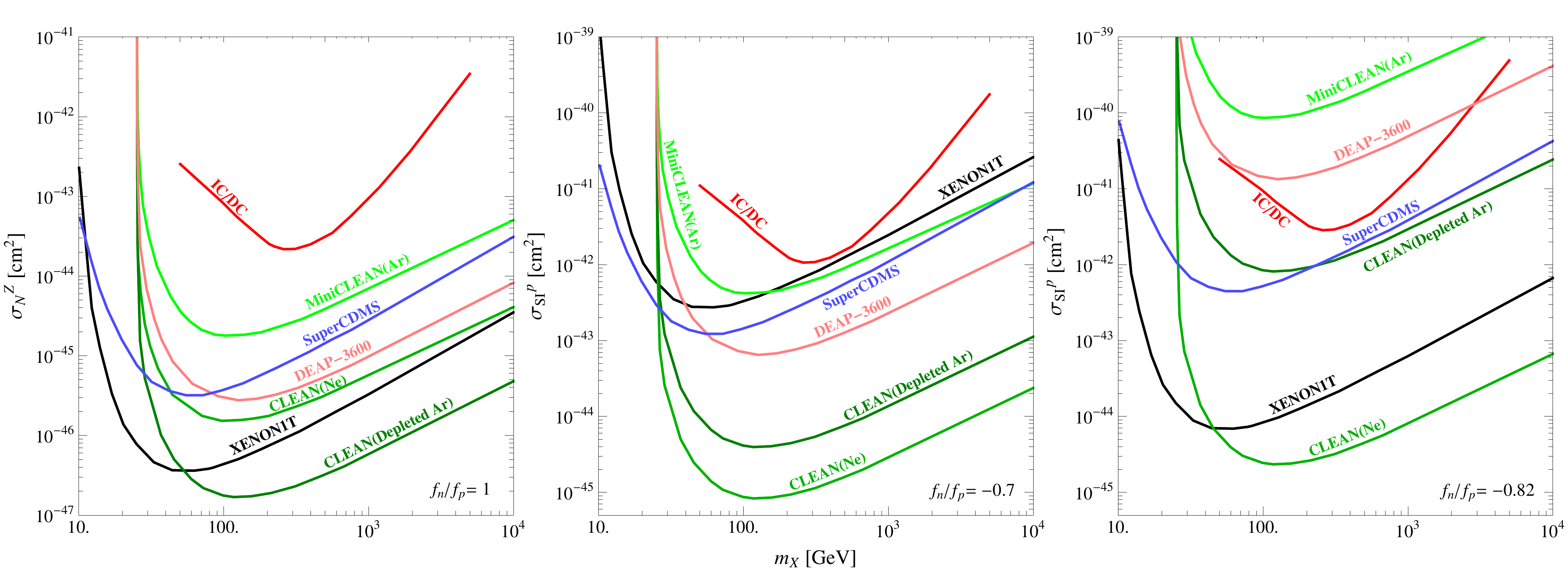}
\vspace*{-.1in}
\caption{\label{fig:sigmaSIoptimalfuture}
Sensitivity to $\sigmaSI^p$ for  $f_n/f_p=1$ (left panel),
$f_n/f_p=-0.7$ (center panel) and $f_n/f_p=-0.82$ (right panel) for
IC/DC with 180 days of data, and for other labelled experiments (see text).
}
\end{figure}

The sensitivity of IceCube/DeepCore to IVDM is determined from its
sensitivity to the spin-dependent scattering cross-section,\cite{Heros:2010ss} scaled by
the ratio of capture rate for IVDM to that for purely spin-dependent scattering.
The capture rate for IVDM can be computed for any choice of $f_n / f_p$ by
correctly accounting for the cross-section for scattering off any element in
the sun.  These bounds are shown in fig.~\ref{fig:sigmaSIoptimalfuture} (see ref.~\refcite{arXiv:1108.0518}),
along with expected sensitivities from XENON1T,\cite{bib:xenon100}
SuperCDMS (100 kg target mass),\cite{Brink:2005ej}
MiniCLEAN, DEAP-3600, and CLEAN (neon or depleted argon).\cite{DEAPCLEAN}

We see that for $f_n / f_p \sim -0.7$, IceCube/DeepCore can, with 180 days of
data, have sensitivity exceeding all current detectors for $m_X > 50~\gev$,
and even rivals the sensitivity achievable with XENON1T.\cite{arXiv:1108.0518}
Similarly, for $f_n / f_p \sim -0.82$, a value for which the sensitivity of argon detectors
is maximally suppressed, IceCube/DeepCore's sensitivity would exceed
MiniCLEAN and DEAP-3600, and would be comparable to that of CLEAN (with an argon
target).

\section{Conclusions}

With exciting potential evidence for dark matter arising from DAMA, CoGeNT and
CRESST, along with improving exclusion bounds from experiments like
XENON100, there is a renewed focus on detailed comparison of results from different
detectors.  Most often, this comparison is made under the assumption that dark
matter interactions are isospin-invariant.  Relaxation of
this assumption can have dramatic effects on the
consistency (or tension) between different data sets.

In particular, IVDM can relieve the tension between the exclusion contours of xenon-based
experiments, and signal regions of other experiments.  But the fortuitous presence of
many xenon isotopes with significant abundance ensures that xenon-based experiments will,
with more data, be able to probe IVDM models which could match the low-mass data.

Several experiments with low-neutron targets, such as COUPP, may soon provide
sensitivities to IVDM complementary to that available to xenon-based experiments.
Neutrino-based indirect searches at Super-Kamionkande and KamLAND can potentially probe
low-mass IVDM with data already
taken.  For higher mass dark matter, IceCube/DeepCore will, with 180 days of data,
have a sensitivity to some regions of IVDM parameter-space which exceeds all current
detectors, and is comparable even to many planned experiments.

A variety of new data will soon come from different detectors,
and IVDM may be an important piece in interpreting this data.  Models of IVDM can be
constrained by gamma-ray\cite{Kumar:2011dr} and collider searches,\cite{Rajaraman:2011wf}
and new data will provide further tests.

\section*{Acknowledgments}

JK gratefully acknowledges, J.~L.~Feng, Y.~Gao, J.~G.~Learned, D.~Marfatia, M.~Sakai, D.~Sanford,
and S.~Smith, who collaborated on the work discussed here.  JK is supported by
DOE grant DE-FG02-04ER41291.


\end{document}